# Universal Free Energy Landscape Produces Efficient and Reversible Electron Bifurcation

**Authors:** J. L. Yuly[1], P. Zhang*[2], C.E. Lubner[3], J.W. Peters[4], and D.N. Beratan*[1,2,5]

[1]Department of Physics, Duke University, Durham, NC  27708
[2]Department of Chemistry, Duke University, Durham, NC  27708
[3]National Renewable Energy Laboratory, Golden, Colorado 80401
[4]Institute of Biological Chemistry, Washington State University, Pullman, Washington 99163
[5]Department of Biochemistry, Duke University, Durham, NC  27710
*Correspondence to:  peng.zhang@duke.edu and david.beratan@duke.edu

**Abstract:** For decades, it was unknown how electron bifurcating systems in Nature prevented energy-wasting short-circuiting reactions that have large driving forces, so synthetic electron bifurcating molecular machines could not be designed and built. The underpinning free energy landscapes for electron bifurcation were also enigmatic. We predict that a simple and universal free energy landscape enables electron bifurcation, and we show that it enables high-efficiency bifurcation with limited short-circuiting (the EB-scheme). The landscape relies on steep free energy slopes in the two redox branches to insulate against short-circuiting using an electron occupancy blockade effect, without relying on nuanced changes in the microscopic rate constants for the short-circuiting reactions. The EB-scheme thus unifies a body of observations on biological catalysis and energy conversion, and the scheme provides a blueprint to guide future campaigns to establish synthetic electron bifurcating machines.

**Significance:** A central challenge faced by molecular machines is to prevent energy wasting short-circuiting (slippage) reactions. Electron bifurcation is an efficient and reversible redox reaction at the heart of key bioenergetic and biocatalytic reaction pathways used in Nature. Electron bifurcation oxidizes a two-electron donor, using the electrons to reduce cofactors on two separate electron-transfer redox chains. The coupling of these redox reactions allows one of the electrons to move thermodynamically uphill, leveraging the downhill flow of the other electron. Thus, electron bifurcation may generate strong oxidants or reductants with minimal free energy loss (i.e., reversibly). Not surprisingly, life harnesses electron bifurcation in biochemical pathways that perform challenging chemical reactions, including proton translocation across membranes, nitrogen fixation, and $CO_2$ reduction. We predict that there is one universal free energy landscape that supports efficient electron bifurcation reactions.

Living systems depend crucially on the efficient interconversion of energy at the molecular scale. Electron bifurcation was recognized by Peter Mitchell as being a key element of the Q-cycle in mitochondria (1), but it now describes a broader class of chemical reactions - presently found only in biology - that oxidize a two-electron donor and reduce two spatially separated one-electron acceptors (2-5). One of the electron transfer reactions from the bifurcating species can proceed thermodynamically "uphill" with respect to the two-electron (midpoint) reduction potential of the electron-bifurcating donor, provided that the other electron proceeds sufficiently downhill for the reaction to be spontaneous overall. Thus, electron bifurcation, or its reverse reaction known as electron confurcation, can occur spontaneously. The near free energy conserving nature of electron bifurcation is the source of its efficiency and novelty; this coupling of "downhill" and "uphill" electron transfers is astonishingly useful. For example, electron bifurcation is used in the Q-cycle of respiration (6) and photosynthesis (7), and to generate low-potential equivalents for $CO_2$ reduction in methanogenesis (8, 9), nitrogen fixation by nitrogenase (10), hydrogen production by hydrogenases (11), and more (4, 12-16). This use of electron bifurcation by Nature to achieve difficult chemical transformations highlights its fundamental place in the toolbox of biological energy transduction (2, 5, 17), and makes electron bifurcation an attractive candidate for biomimetic energy schemes that require the production of highly reducing or oxidizing species (3, 5, 18).

**Short circuiting limits electron bifurcation efficiency**. The process of electron bifurcation is illustrated in Fig. 1 A. First, a two-electron donor (D), with a mean reduction potential in the middle of the physiological window, donates its electrons to the electron bifurcating enzyme. The electrons reach the electron bifurcating cofactor (B), which sends one electron into a low potential hopping pathway and one into a high potential hopping pathway (cofactor chains L and H, respectively). These paths each terminate at electron accepting substrates, one at high ($A_H$) and the other at low ($A_L$) reduction potential. In the reverse (confurcating) reaction, one electron flows from $A_H$ and another electron from $A_L$ to doubly reduce the bifurcating species B, which then performs a two-electron reduction of D. For efficient electron bifurcation to occur, one electron must proceed through the low-potential branch for every electron that flows through the high potential branch. Efficient electron confurcation requires that every electron flowing from the high-potential substrate $A_H$ to species B must be matched with an electron from the low potential substrate $A_L$ to reduce B. In most electron bifurcating systems, B is either a quinone or a flavin (4), although transition metal complexes may also bifurcate electrons (19). The L- and H-cofactors typically include hemes, iron-sulfur clusters and/or non-bifurcating quinones and flavins (4, 19).

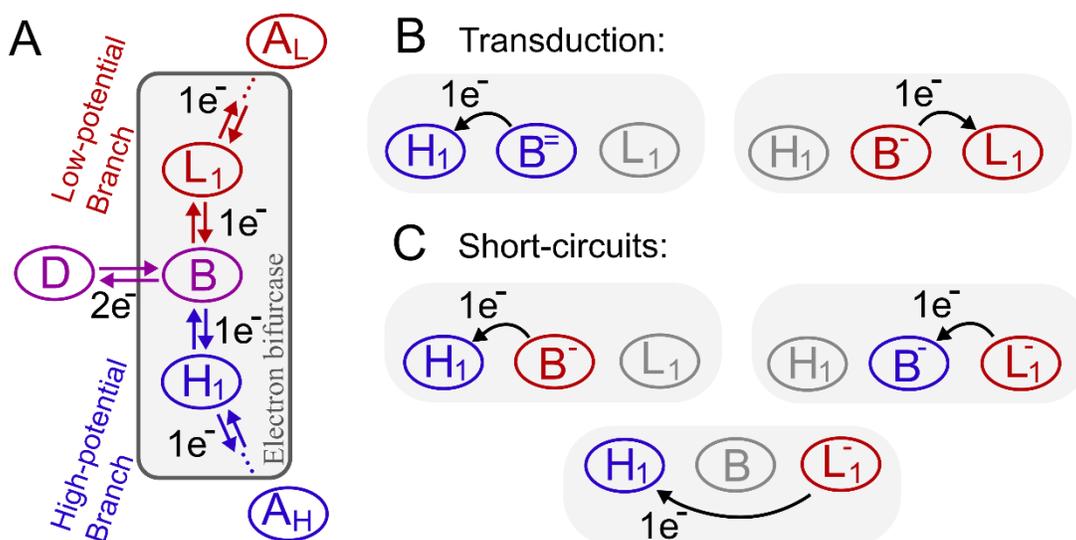

**Fig. 1. Electron bifurcation and short-circuit pathways.** (A) The kinetic network underpinning an electron bifurcating enzyme, and the redox reactions that may take place at the bifurcating site B. There may be additional cofactors in either chain (for instance $L_3$ or $H_4$, not explicitly shown). Red indicates the low redox potential (high energy) path, while blue indicates the high-potential (low energy) path. Purple indicates the two-electron bifurcating species B, and the two-electron source/sink D. Electron transfer from the electron bifurcating site can result in energy transduction (B), but energy wasting short-circuit reactions (C) also occur between the same cofactors.

Nature's electron bifurcation machinery has proven difficult to imitate, and no synthetic molecular machine has been built that carries out high efficiency electron bifurcation. The obstacle to realizing efficient electron bifurcation arises from the short-circuiting reactions intrinsic to the bifurcating network, indicated in Fig. 1 C (5, 20, 21). Short-circuit electron transfer reactions occur when an electron flows from the $B^-$ intermediate to the high-potential acceptor $A_H$, or when electrons individually flow from the low-potential (high-energy) branch to reduce $B^-$. In addition, direct tunneling from $L_1$ to $H_1$ is possible, although the tunneling distance is very large in known electron bifurcating enzymes (~ 20 Å or more) (21, 22), substantially slowing this short-circuit reaction.

The Q-cycle was the first electron bifurcation reaction that was found to be reversible on relevant physiological timescales (20). Since the tunneling distances for short-circuit transfers (Fig. 1 C) are the same as for productive transfers, the rate constants for the productive electron transfers are expected to be similar to those for the short-circuit electron transfers (21). To prevent short-circuiting, "gating mechanisms" were proposed to suppress short-circuiting reactions, including concerted two-electron transfer (21), conformational gating (5, 23), "spring-loading" of the Rieske iron-sulfur protein (24), Coulombic interactions (25), and other possible mechanisms termed double redox gating (20, 21). However, after almost 20 years of searching, no experimental "smoking gun" in support of these gating mechanisms has been found. For example, it is understood that conformational motion of the Rieske iron-sulfur protein is required to explain how electrons tunnel through the high-potential branch. But this conformational motion does not itself serve as a gating mechanism (to suppress short-circuiting electron transfer rate constants) because the reactions operate under near reversible conditions (5, 6, 21). Indeed, there is no consensus on how the Q-cycle accomplishes reversible operation with such high efficiency.

In addition to the quinone-based Q-cycle complexes, other novel flavin-based electron bifurcating enzymes were discovered in the last decade (4, 5, 9, 26, 27). Many (if not all) of these flavin-based electron bifurcating enzymes are also reversible (4), and many are not membrane bound (19, 22); others seem to lack significant conformational flexibility (3, 5, 19). Short-circuiting electron transfer also creates a challenge to flavin-based electron bifurcating enzymes (5), and how these bifurcating flavoenzymes avoid short-circuiting, while maintaining reversibility, is unknown.

**Introducing a thermodynamic landscape to enable efficient electron bifurcation.** The analysis presented here indicates that a universal mechanism of high efficiency bifurcation is used by all electron-bifurcating enzymes. We find that the secret to avoiding slippage (short circuiting electron transfer) in electron bifurcation reactions lies in the steep free energy (reduction potential) landscapes of the spatially separated high- and low-potential branches, which is considered to be an enigmatic (but conserved) feature of electron bifurcating enzymes (4, 5, 28). This landscape has a form similar to the redox potential landscapes in photosynthesis (29), although the mechanism for electron bifurcation is drastically different from that of photosynthesis.

In Nature, steep free energy landscapes are not unique to electron bifurcation. Photosynthesis uses steep landscapes to prevent charge recombination and to induce high-yield electron transfer following photoexcitation (29). Fig. 2 shows nine possible free energy landscapes for electron bifurcation, discussed in detail below. Only one landscape, indicated in panel G of Fig. 2, supports efficient electron bifurcation by suppressing short-circuiting (*vide infra*). Without the EB-scheme design principle, successful synthetic electron bifurcation (i.e., the equal and reversible yield of the high- and low-potential redox products) seems tremendously difficult to accomplish. This free energy design principle, described and analyzed in detail below, explains how Nature elegantly skirts a major obstacle (short-circuiting reactions) to producing high-value redox species.

**Candidate free energy landscapes for electron bifurcation.** There are three main ways that the thermodynamic landscape may influence electron transfer rates in an oxidoreductase. First, electron transfer rate constants in proteins are determined by tunneling pathways and distances between cofactors, reorganization energies, and thermodynamic driving forces (30). Thus, the reduction potential landscapes of the electron bifurcation branches, the cofactor placement, and the protein structure (31) determine the productive and short-circuit electron transfer rate constants. Second, the thermodynamic landscape establishes steady-state populations for each possible redox state. Indeed, these steady-state populations determine the effective activation free energies for short-circuiting electron transfer (*vide infra*). Third, the free energy difference between initial and final catalytic states determines the catalytic driving force (and hence whether the reaction runs in the forward or reverse direction). The overall driving force for electron bifurcation is

$$\Delta G_{bifur} = 2FE_D - FE_{A_L} - FE_{A_H} \qquad [\mathbf{1}]$$

where $E_D$, $E_{A_L}$, and $E_{A_H}$ are the (midpoint) reduction potentials of the D, A$_L$, and A$_H$ substrates, respectively, and $F$ is Faraday's constant. For electron bifurcation to be spontaneous, $\Delta G_{bifur} < 0$.

Nine possible free energy landscapes for electron bifurcation are categorized in Fig. 2. Landscapes A, B, and C have $\Delta G_{bifurc} \ll 0$ and hence are not reversible, only operating in the electron bifurcation direction. Landscapes D, E, and F have $\Delta G_{bifurc} \gg 0$ and only operate in the electron confurcating direction. Thus, only landscapes G, H, and I, with $\Delta G_{bifurc} \approx 0$, are suited for reversible electron bifurcation/confurcation. To drive catalysis in the electron bifurcating (confurcating) direction with these landscapes, one would simply tune the reduction potentials of the terminal substrates to tilt the free energy balance slightly (Eq **1**) (via reactant concentrations or the transmembrane potential for membrane-bound proteins). The reversibility of electron bifurcation is the source of its energetic efficiency (3, 5, 32).

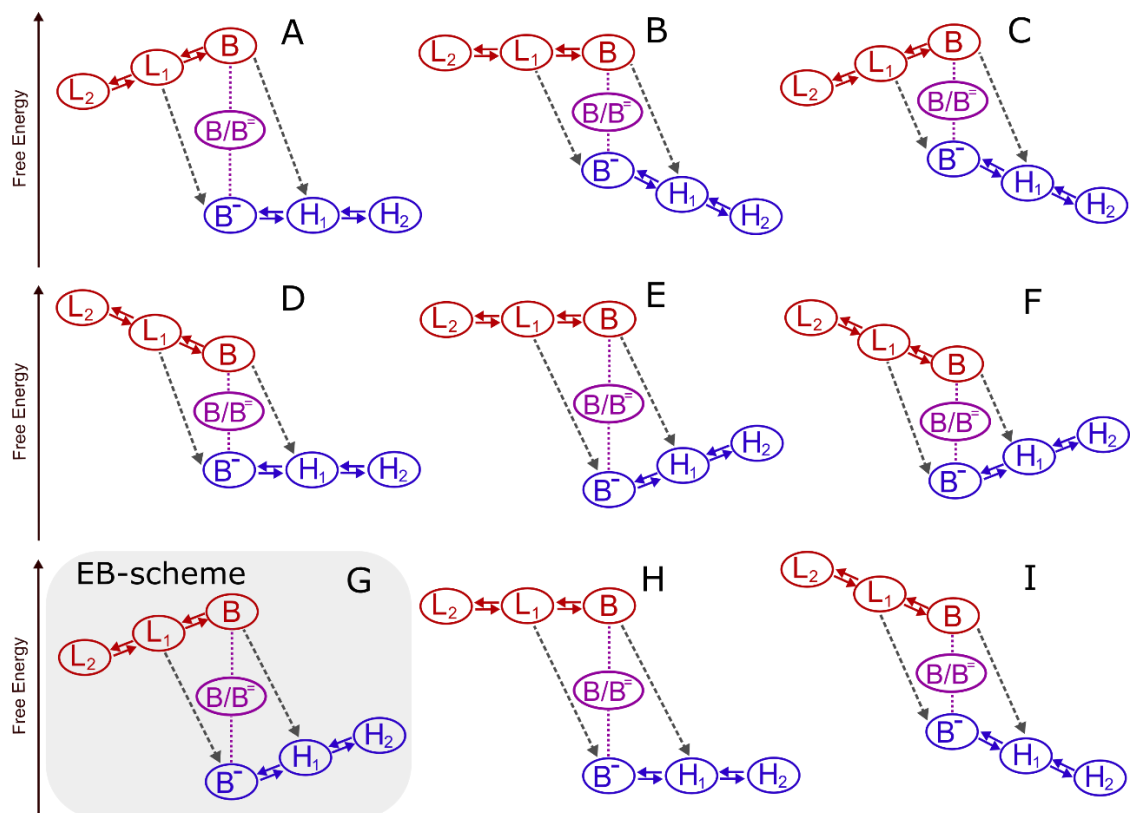

**Fig. 2. Candidate free energy (reduction potential) landscapes for energy conserving electron bifurcation.** Solid arrows represent productive electron transfer steps, and grey dashed arrows represent short-circuit energy dissipating steps. The purple ovals represent the positions of the two-electron (midpoint) reduction potential of the bifurcating species. Landscapes (A-F) are not free energy conserving (i.e. they produce irreversible electron bifurcation or confurcation) and therefore are not energetically efficient, while landscapes H and I produce short-circuiting (Fig. 3). Only landscape (G) is reversible and avoids short circuits, thanks to the Boltzmann suppression of microstates in which short circuiting can occur.

**The EB-scheme.** Now, we describe how the EB-scheme shown in Fig. 2 G insulates the kinetic network from short circuits, while producing high-efficiency (reversible) electron bifurcation, and we prove this claim numerically in the next section (the other two energy conserving landscapes, illustrated in Figs. 2 H and I, lead to copious short circuiting and are not viable). The slopes of the H and L redox branches in Fig. 2G cause electrons to pile up near B in the low-energy branch (blue), and holes in the high-energy branch (red) near B. Since the one-electron cofactors cannot accept a second electron at relevant potentials and must be in the reduced state to donate an electron, the EB-scheme insulates the enzyme against short-circuiting by an electron occupancy blockade effect, despite having large short-circuiting rate constants. For an energy wasting short-circuiting reaction to occur, a hole must occupy the low-energy branch (blue) and an electron must occupy the high-energy branch. Taken together, these processes create a very large free energy barrier for short-circuiting. That is, the EB-scheme is protected against short-circuits by Boltzmann occupancy factors, so the enzyme will rarely enter a state where short-circuits can occur. For productive electron bifurcation (confurcation) to occur, only a hole (electron) must move down (up) the low (high) energy branch, so the productive transfers have a much smaller free energy of activation to overcome. This occupancy effect, arising from the EB-scheme landscape, can lead to highly efficient partitioning of electrons into the high and low potential branches. The viable EB-scheme (Fig. 2G), examined in detail here, uses crossed potentials at the bifurcating site B, but we have not examined whether crossed potentials are a requirement for effective electron bifurcation; the role of crossed potentials in electron bifurcation was discussed recently (3, 5, 12, 32). Next, we show how these principles emerge quantitatively from a kinetic model for electron bifurcation that explicitly describes the electron flux through the kinetic network, explicitly including cofactor occupancy effects.

**Many-state, many-electron kinetics of electron bifurcation.** Attempts were made in earlier studies to model the kinetics of electron bifurcation, and those studies succeed in describing many features of the kinetics. However, some of the previous models are not reversible (3, 21) and, as such, are inconsistent with the known reversibility of biological electron bifurcation. Other models restrict the number of tunneling electrons in the enzyme to just two (33) (inconsistent with access to pools of one- and two-electron redox substrates) or use rate constants that are physically unmotivated (34, 35), including ad hoc turning off of short-circuit reactions (36, 37). The scheme described here avoids these unnatural constraints and treats productive electron transfers (Fig 1 B) on the same footing as short-circuits (Fig 1c), allowing electrons to tunnel freely with rate constants estimated using non-adiabatic electron transfer theories with appropriate Marcus factors (30, 38), but only when a mobile electron resides on the donor, and a hole on the acceptor (i.e., we explicitly track the occupancies of all redox-active species). The substrates $D^=$, $A_H$, and $A_L$ were modeled as electron reservoirs, which release and accept electrons at the reduction potential of the substrate, with adjustable rate constants that were tuned so that they are not rate limiting (that is, the intrinsic kinetics of the electron bifurcating enzyme are assumed rate limiting). Electrons move two-at-a-time in one kinetic step into B. For the Q-cycle, this describes quinone diffusion into the $Q_0$ site. Details of the kinetics model appear in the *SI Appendix*.

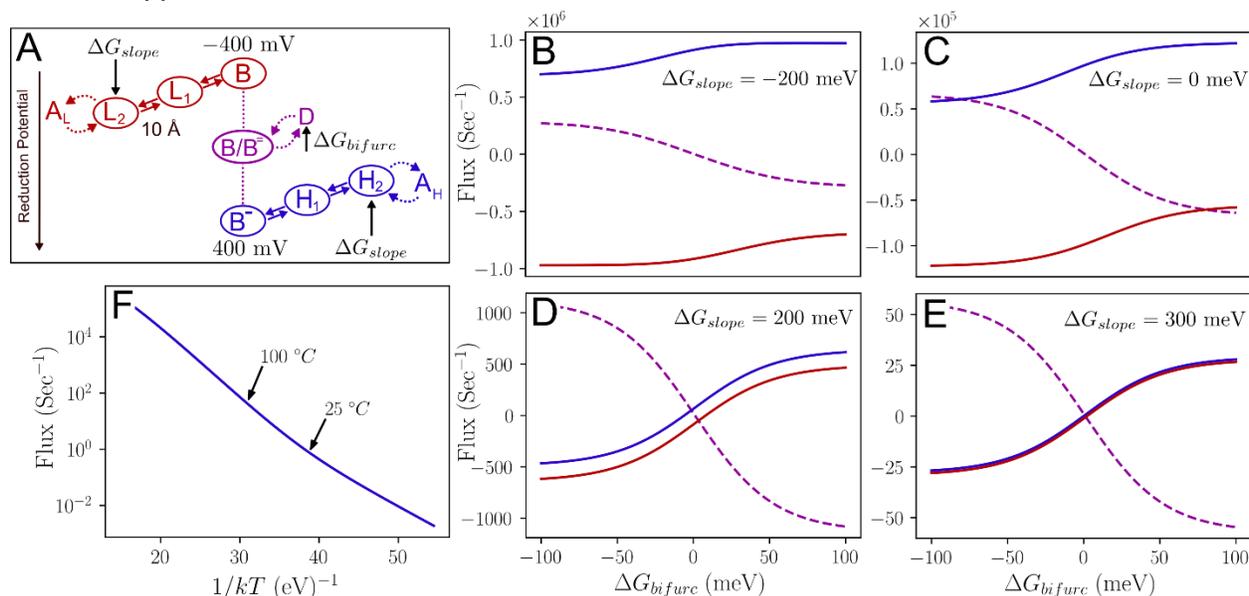

**Fig. 3. Model electron bifurcating enzyme and turnover kinetics as a function of driving force.** (A) Model electron bifurcating enzyme with cofactors spaced by 10 Å, and with the free energy landscapes of the branches characterized by $\Delta G_{slope}$, the energy to move an electron from the bifurcating site to the terminal substrate. The one-electron substrates were modeled as electron reservoirs with reduction potential equal to the terminal cofactor potentials on the two branches. (B-C) net electron fluxes into the $A_H$ (blue), D (purple) and $A_L$ (red) reservoirs as a function of $\Delta G_{bifurc}$. As the slope of the branches changes from (B) negative to (C) flat to (D) positive to (E) steeply positive, the dynamics change continuously from short-circuit dominated (B-C) to electron bifurcation/confurcation dominated with significant short-circuits (D), to transduction with negligible short-circuiting (E). The magnitudes of all fluxes significantly drop with increasing $\Delta G_{slope}$, reflecting the high energy barriers for electron flux between substrates in the EB-scheme. For $\Delta G_{slope} = 300\ meV$ and $\Delta G_{bifurc} = 0$, the inverse-temperature dependence of the short-circuit fluxes (F) has two clear linear regimes, which indicates a thermally activated mechanism (note the log-linear plot).

For each of the three free energy conserving schemes (Fig. 2 G-I) for electron bifurcation, we implemented a minimalistic kinetic model, *mutatis mutandis*, for electron bifurcating enzymes. The model (Fig. 3 A), and the resulting kinetics at steady state (Fig. 3 B-F), are shown in Fig 3. The $B/B^-$ and $B^-/B^=$ standard reduction potentials were set to -400 mV and 400 mV, respectively, and the nearest-neighbor distance between cofactors was set to 10 Å (next-to-nearest distance of 20 Å, etc.). Nature's electron bifurcation systems vary these parameters, but the chosen values are typical (4). While the efficiency and turnover time can be tuned by changing these parameters (Fig. S1), energy dissipating rapid short-circuiting ($\approx 10^5$/sec) as in Fig. 3 B-C is never observed when the EB-scheme is present. Nearly perfect one-to-one partitioning of electrons to the high and low potential substrates with full reversibility can be accomplished without requiring a gating mechanism (Fig. 3 E).

We explored the short-circuit behavior of the landscapes in Fig. 2 G-I as a function of the driving force $\Delta G_{bifurc}$ with this kinetic model. For the landscape of Fig. 2 I, the electron flux away from $A_L$ and into $A_H$ is large (~ $10^6$ electrons/sec), reflecting short-circuit dominated kinetics. For landscape H, the short-circuiting flux is still large (~ $10^5$ electrons/sec). Only when the slope of the branches follows landscape G (the EB-scheme), do the electron fluxes into $A_H$ and $A_L$ have the same sign, reflecting electron bifurcation (confurcation) dominated kinetics when the overall driving force $\Delta G_{bifurc}$ is negative (positive). Any difference between the $A_H$ and $A_L$ oxidation/reduction rates (separation between the red and blue curves) reflects short-circuiting behavior, so the near superposition of the curves in Fig. 3 E indicates very low short-circuit currents.

**The EB-scheme suppresses electron short-circuiting.** When the magnitude of the energetic slopes of the two EB-scheme redox pathways is increased (Fig. 3 E), the short-circuiting flux shrinks compared to the electron bifurcating/confurcating turnover rates, as reflected in the negligible difference between the electron fluxes into/out of the $A_L$ and $A_H$ reservoirs. Using the EB-scheme, electron bifurcation can achieve high efficiency (equal partitioning of electrons into the $A_L$ and $A_H$ reservoirs), at the cost of turnover speed and reducing power of the low-potential acceptor $A_L$. Presumably, electron bifurcating enzymes in Nature evolved to balance these tradeoffs, insulating against short-circuits while enabling catalysis to proceed with sufficient speed to meet physiological demands. Importantly, alternate gating mechanisms are not required for reversible and efficient electron bifurcation in the EB-scheme. In fact, electron bifurcation and confurcation emerge naturally from the kinetic network (Fig. 2 G and 3 A) at steady state, but only when the EB-scheme is employed. Our model does not unnaturally privilege productive electron transfers over short-circuits in any way. Indeed, short-circuit electron transfers are successfully insulated in the EB-scheme, even when the short-circuit rate constants are set orders of magnitude faster than the productive electron transfers, due to the cofactor occupancy blockade effects (Fig. S1 D)!

When short-circuit fluxes are small (i.e., as occurs in the EB-scheme), the high- and low-potential redox branches quickly reach approximate chemical equilibrium with themselves, despite being out-of-equilibrium with the other branch (6) (i.e., quasi-equilibrium). Thus, the short-circuit fluxes are thermally activated. Fig. 3 F shows the short-circuit flux into the high potential $A_H$ reservoir when $\Delta G_{bifurc} = 0$ (the electron bifurcating enzymes is "idling") as a function of temperature, where two distinct linear regimes are observed at low and high temperature, which indicates a thermally activated tunneling mechanism for the short-circuits (this linear behavior is analyzed in detail in the *SI Appendix)*. The high-temperature regime is dominated by $B^-$ mediated short circuits, which are fast but have a large thermal activation. The low-temperature regime is dominated by the $L_1$ to $H_1$ short-circuit, which is slower but has a smaller thermal activation energy, allowing this short-circuit to dominate at low temperatures.

The energetic landscapes of electron bifurcation have been proposed to be important many times before (e.g., see refs. (3-5, 28, 32, 36)), but the special and universal nature of the EB scheme to nearly eliminate short-circuits and remain fully reversible has not been shown previously. This is because a minimalistic model must include the potent combination of: A) reversibility (20, 21), B) explicit tracking of the entire enzyme's redox state (not just the average state of each cofactor) (35, 37), C) three explicit electron reservoirs that are each free to exchange electrons in the branches at each reservoir's chemical potential, and D) the explicit modeling of the energetic slopes along the entire length of the high- and low- potential branches, not just the cofactors near the bifurcating site (6). In fact, to our knowledge, the model described here is the first to explicitly show a reverse electron flux with negligible short-circuiting when the driving force, $\Delta G_{bifurc}$, is reversed.

While reversibility, electron blockading, and explicit reservoirs are crucial to capture efficient electron bifurcation, combining all three into a tractable kinetic model is not simple because the number of differential equations governing the kinetics grows exponentially with the number of cofactors (35, 37). To construct the very large model that underpins Fig. 3, we procedurally generated the equations governing the dynamics (see *SI Appendix*). In fact, our model is similar to that found in refs (35, 37), except that we answer quantitatively the apparently central question, namely why electron bifurcating enzymes never use any of the landscapes in Fig. 2, aside from landscape G. Understanding precisely how landscape G insulates against short-circuits allows us to make the strong prediction that landscape G of Fig. 2 (the EB scheme) is universal in electron bifurcation, and that this scheme is key for the design of synthetic electron bifurcating systems (see the next section).

Interestingly, the EB-scheme privileged landscape for electron bifurcation follows a free energy profile that is similar to the steep slopes in reduction potentials that are found in the Z-scheme of photosynthesis (29), However, the mechanism of electron flow in bifurcating enzymes is drastically different, as a consequence of reversibility of electron bifurcation reactions, in contrast to strongly driven photosynthetic reactions.

Electron bifurcating enzymes can surely exhibit complexities that are not captured in our model. For example, proton-coupled electron transfer (5, 6), two-electron cofactors (flavins or quinones) in the H and L branches (22), conformational changes (23, 39), and electron transfer between electron bifurcating monomers (40) may all add kinetic richness. In fact, conformational motion in the Q-cycle is understood to be required for electrons to reach the high-potential cytochrome $c_1$, which is too far away for direct electron tunneling from the electron-bifurcating $Q_o$ site (33). However, none of these specific features interfere with the essential short-circuit insulating nature of the conserved and predicted universal EB-scheme.

**Short-circuiting in the Q-cycle**. A fully-detailed kinetic model of the Q-cycle is beyond the scope of this study, but a simplified model is sufficient to account for the primary cause behind the short-circuit insulation in the Q-cycle. Our model (Fig. 4A) uses distances and energetics suggested by experiment and indicates that the EB energy landscape explains most of the short-circuit insulation in the Q-cycle (see also *SI Appendix Supplementary Text*). Cofactor reduction potentials were measured previously (28), and the tunneling distance values were used in previous studies (33). The first electron transfer from $Q_o$ to ISP (ISP = iron-sulfur protein) is proton-coupled and rate limiting (6). This was modeled by setting an effective electron tunneling distance, which was tuned until the overall steady-state turnover was ~ 50/sec, placing the model in quantitative agreement with experimental steady-state turnover rates (6, 40). This fitting procedure forced the SQ→ISP (SQ = semiquinone) short-circuit rate constant to be favored over the productive HQ→ISP (HQ = hydroquinone) rate constant by several orders of magnitude. Even with this preference for a short-circuit rate constant over a productive one, short-circuits were still successfully insulated (Fig. 4 B-C). The motion of the ISP was not explicitly modeled, but was assumed to be sufficiently fast so that electrons can tunnel directly to cytochrome $c_1$ once ISP is reduced. The $Q_i$ site was modeled as a one-electron reservoir, since the two one-electron reduction potentials of ubiquinone at the $Q_i$ site are similar (41).

A few specific experiments have been interpreted as indicating a need for alternative gating mechanisms in the Q-cycle to assure its efficient function. For instance, in the Q-cycle of cytochrome $bc_1$, the inhibitor antimycin A (which prevents electrons from leaving the low potential branch) is known to decrease the overall steady-state turnover by a factor of about 30 (6, 40). Gating mechanisms were proposed to explain this slowdown of the redox flux with a compromised L-branch (20, 21, 33, 37). Our simplified model of the Q-cycle using experimental parameters (Fig 4) shows that the EB-scheme insulates against short circuits in that system. Not surprisingly, slower turnover is not observed in our simplified kinetics model with an inhibited L-branch as compared to the uninhibited case (Fig S3), which suggests that additional features not captured in our model play a role in the Q-cycle.

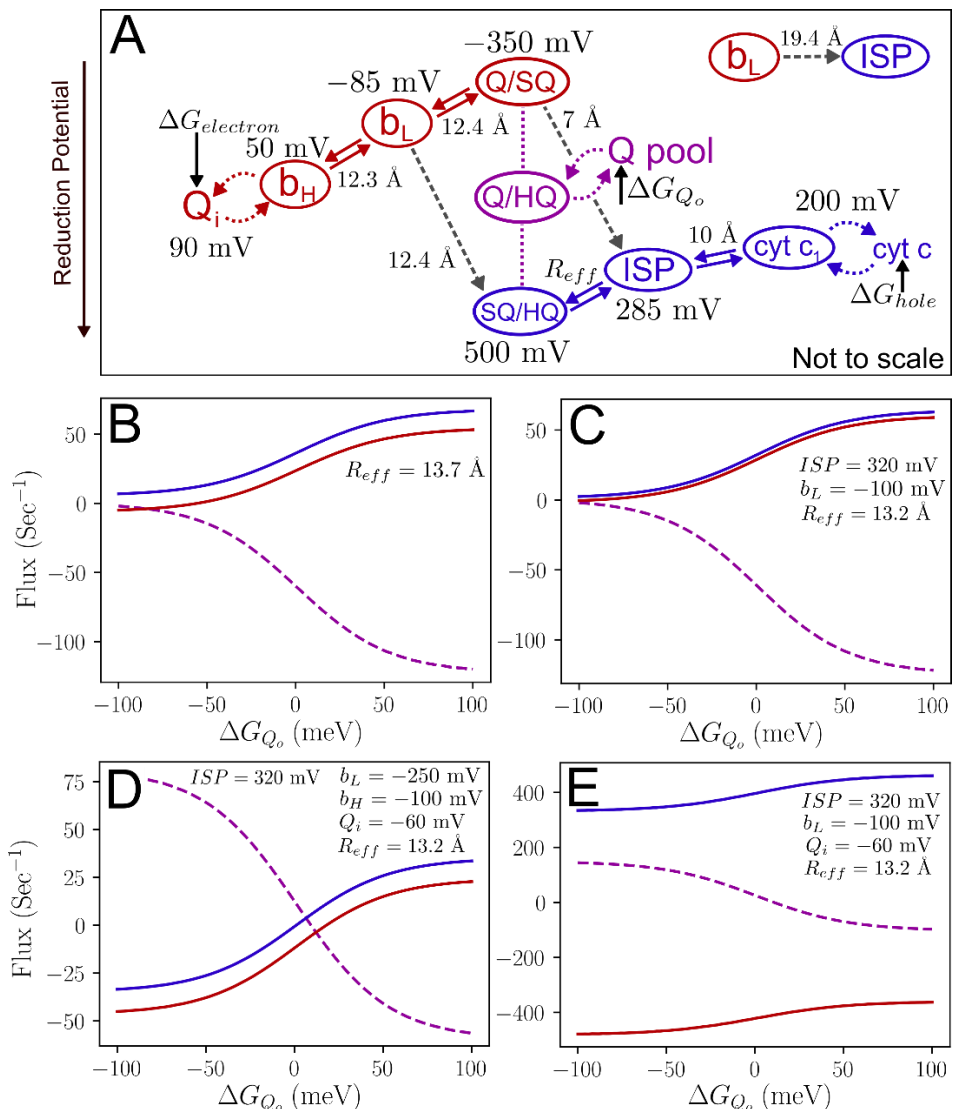

**Fig 4. Short-circuit insulation in the Q-cycle (complex III) arising from the EB-scheme.** Using our multi-electron kinetic model (*SI Appendix*), we built a simplified model for the Q-cycle using previously published reduction potentials and tunneling distances (28, 33). Despite these simplifications, the EB-scheme observed in the measured reductions potentials seems (B) effective at insulating against short-circuits. With minor changes to the reduction potentials (designed to increase $\Delta G_{electron}$ and $\Delta G_{hole}$ of the activation process shown in Fig. S2 D) that are likely within the range of experimental uncertainty (C), the EB-scheme of the Q-cycle provides the preponderance of insulation against short-circuiting. In (B-C), no confurcation appears for the values of $\Delta G_{Q_o}$ shown since the reduction potentials cited were measured in the absence of the membrane potential (28), by which energy is ultimately conserved in the Q-cycle. The influence of the membrane potential on all of the cofactor reduction potentials (and hence the EB-scheme) is unknown. We present two possible cases, chosen to reflect the range of possible impacts of the membrane potential on the L-branch cofactor reduction potentials. In (D) the reduction

potentials of the low-potential branch all decrease by 150 mV. In (E) the reduction potential of the $Q_i$ site alone decreases by 150 mV. In the case of (D), the EB landscape may be sufficiently preserved to insulate from short-circuits. In (E), the EB landscape is significantly disrupted, as the energy required to move an electron to cytochrome $b_L$ (in Eq **S8**) in order to initiate short-circuiting is negligible. This disruption of the landscape turns on short-circuiting, but may not reflect the reality of cytochrome $bc_1$ in the presence of a membrane potential.

Molecular features behind the observed difference between uninhibited and L-branch inhibited kinetics in the Q-cycle may include subtle structural changes resulting in tunneling distance changes of about 3 Å or less between the $Q_o$ site and its iron-sulfur cofactor partner (Fig. S4), or subtle electrostatic interactions between the low-potential branch and the $Q_o$ site (*SI Appendix*, these are likely not the only possible explanations), rather than by a gating mechanism *per se.* Because the measured change in steady-state turnover in the presence of inhibitors and L-branch cofactor knockouts (20, 40) is so subtle (about a factor of 30 less (40)), these and other mechanisms will be difficult to identify uniquely (See *SI Appendix* for extended discussion).

The effect of the EB-scheme in preventing short-circuits is orders of magnitude larger than the observed difference between the L-branch inhibited and non-inhibited turnover. Specifically, the rate constants for short-circuit electron transfers are $\sim 10^9$/sec (35), which must be defeated. In the absence of additional assistance from protein gating, the EB-scheme will reduce the flux to $\sim 10^2$/sec, and these values will be further reduced if the L-branch is not inhibited (see Fig. 3 and Fig. S3), supporting the central role played by the EB-landscape in defeating short circuits. Only about one additional order of magnitude is needed to bring this turnover rate to the observed L-branch inhibited turnover rates ($\sim 10^0$/sec) (40), which indicates that the efficiency gained by such a mechanism is less than 1% of the gain produced by the EB-scheme (measuring efficiency with short-circuit rates). Thus, the EB-scheme explains most of the short-circuit insulation in the Q-cycle in cytochrome $bc_1$.

**Natural electron bifurcation: exploiting the EB-scheme.** The important lesson learned is that any additional mechanisms in natural electron bifurcating enzymes, beyond the EB-scheme, are not the key features that underpin the short-circuit insulation in electron bifurcating systems, including the Q-cycle. There is a tremendous difference between a gating mechanism that changes a rate constant by nine orders of magnitude (which is required to insulate against short-circuits without the EB-scheme) and a mechanism that intrinsically prevents that kinetic pathway from ever being accessed by the system under normal operating conditions (this is how the EB-scheme works, see Fig S2). Other subtle features and mechanisms might shave off the last order of magnitude or two of short-circuiting flux when the L-branch is inhibited, or even permit some short-circuiting and serve as "release valves" that can reroute electrons to add robustness to biochemical pathways. For instance, certain photosynthetic bacteria were found to be able to grow with an inhibited Q-cycle (42). These organisms required a short-circuit flux across a Q-cycle to grow. Importantly, our analysis does not disentangle subtle features of electron bifurcating enzymes, which likely differ from system to system. We do, however, propose that the EB-scheme is sufficient to accomplish robust electron bifurcation, explains the lion's share of short-circuit insulation in known electron bifurcating systems, and may serve as a core design framework for synthetic electron bifurcating systems.

**Synthetic electron bifurcation: exploiting the EB-scheme.** The EB-scheme enables reversible electron bifurcation, insulates against wasteful short-circuit reactions, and thus appears to remove the two primary roadblocks that prevent the design and synthesis of electron bifurcating molecular machines (5). A robust and general scheme to prevent short circuits suggests that synthetic electron bifurcation is not a distant dream.

We envision that EB-schemes (Fig 2 G) may be assembled based on several kinds of molecular architectures. For instance, covalently linked molecular redox species, DNA origami motifs (43), tailored linked quantum dots (44), or even semiconductor nanostructures may serve as possible frameworks in which to realize electron bifurcation. For example, the EB-scheme landscape is found in the band bending in semiconductor n-p junctions (45), which suggests that semiconductors may play a role in synthetic electron bifurcation.

In the EB-scheme, each redox active site must be made to accommodate only one mobile electron at a time and must not be allowed to interact further than its nearest neighbors. For example, if $L_2$ could donate an electron to $H_2$ (Fig. 1), or if $H_1$ could receive several electrons from $B^=$, the EB-scheme would no longer insulate against short circuits (these processes were included in our model, but since the distance between non-neighbor cofactors is at least 20 Å in the model, the corresponding tunneling rate constant is negligibly small). In addition, the terminal electron acceptors, D, $A_L$, and $A_H$, must not exchange electrons directly with each other, or with any of the redox active sites in the scaffold, other than with the terminal branch sites. This level of microscopic control is challenging to realize, and anchoring the $A_L$ and $A_H$ acceptors to the ends of the branches may be acceptable for proof-of-concept experiments. Care must also be taken to avoid short-circuit channels during the $D^=$ to B electron refilling process (short-circuiting during refilling).

Electron bifurcation in Nature allows the reversible reduction of compounds with low reduction potentials, using compounds with much higher (midpoint) reduction potentials, analogous to the function of a voltage amplifier. Understanding the manner of this redox conversion in the warm, wet environment of biology provides inspiration for novel synthetic redox catalysts.

**Acknowledgments:**

This work is supported as part of the Biological Electron Transfer and Catalysis (BETCy) EFRC, an Energy Frontier Research Center funded by the U.S. Department of Energy, Office of Science (DE-SC0012518). This work was authored in part by Alliance for Sustainable Energy, LLC, the manager and operator of the National Renewable Energy Laboratory for the U.S. Department of Energy (DOE) under Contract No. DE-AC36-08GO28308. Funding was provided to C.E.L. by the U.S. Department of Energy Office of Basic Energy Sciences, Division of Chemical Sciences, Geosciences, and Biosciences, Physical Biosciences Program. The U.S. Government and the publisher, by accepting the article for publication, acknowledges that the U.S. Government retains a nonexclusive, paid-up, irrevocable, worldwide license to publish or reproduce the published form of this work, or allow others to do so, for U.S. Government purposes.

We thank Gerrit Schut for helpful discussions.

**Software availability**: The Python scripts used to procedurally generate and to solve the kinetic network equations are posted on GitHub (https://github.com/JYuly/EB_kinetics).

**Author Contributions:**

D.N.B. and P.Z., and J.W.P. conceived the project, J.L.Y. generated the basic hypotheses, J.L.Y., P.Z., and D.N.B. designed and tested the kinetics simulations, J.L.Y. programmed the simulations and generated all figures, J.L.Y and P.Z. drafted the manuscript, and J.L.Y., P.Z., D.N.B., C.L., and J.W.P. edited the manuscript.

# Supplementary Materials for

**Title: Universal Free Energy Landscape Produces Efficient and Reversible Electron Bifurcation**

J. L. Yuly, P. Zhang*, C.E. Lubner, J.W. Peters, and D.N. Beratan*

Correspondence to: peng.zhang@duke.edu and david.beratan@duke.edu

This PDF file includes:

Materials and Methods

Supplementary Text

Figs. S1 to S3

Supplementary references

## Materials and Methods

### Methods:

Electron bifurcases have several cofactors which may be seperately oxidized or reduced. Thus, the redox state of an electron bifurcase may be described by the oxidation state of each cofactor, or the number of mobile electrons present on each cofactor. Such a state may be written as

$$[n_B n_{L_1} n_{L_2} \dots n_{H_1} n_{H_2} \dots] \qquad [S1]$$

where $n_i$ is the number of mobile electrons on cofactor i. For example, the state [21001] corresponds to fully reduced $B^=$, reduced $L_1^-$, oxidized $L_2$, oxidized $H_1$, and reduced $H_2^-$. In general, there are $3 \cdot 2^{N_L} \cdot 2^{N_H}$ possible states, where $N_L$ and $N_H$ are the number of cofactors in the high- and low-potential branches, respectively. This scaling holds assuming each cofactor in the high- and low-potential branches may hold either zero or one mobile electron. Thus, the number of possible states grows exponentially with the number of cofactors, which has previously been pointed out in the context of electron bifurcation in the Q-cycle (1).

These states may be connected by electron transfer rate constants. States are only connected kinetically if the number of electrons is conserved (with the exception of interaction with the reservoirs $A_L$, $A_H$, and $D^=$ discussed below). These rate constants are estimated using a nonadiabatic electron transfer rate expression

$$k_{ET} = \frac{2\pi}{\hbar}\langle|H_{DA}|^2\rangle FC, \qquad [S2]$$

where $\langle|H_{DA}|^2\rangle \approx |H_{DA}^0|^2 e^{-\beta R_{DA}}$. The $FC$ term is the nuclear Frank-Condon factor, which at high temperatures is assumed to have the Marcus form

$$FC = \frac{1}{\sqrt{4\pi\lambda k_B T}} exp[-(\Delta G° + \lambda)^2/(4\lambda k_B T)]. \qquad [S3]$$

The reorganization energy $\lambda$ was estimated at 1.0 eV, a value of 0.1 eV was taken as the contact coupling $|H_{DA}^0|$, the tunneling decay constant $\beta = 1\,\text{Å}^{-1}$, and $k_B T$ (Boltzmann constant times the absolute temperature) was taken at room temperature ($1/k_B T \approx 38.91/eV$). These values are typical (2). In our model bifurcating system (Fig. 3A), the nearest-neighbor distance was set to 10 Å, the next-neighbor distance to 20 Å, etc. These distances are typical of those found in electron bifurcating enzymes (3, 4).

Electron reservoirs were used to model the terminal redox substrates $A_L$, $A_H$, and $D^=$. They connect states (Eq **S1**) that differ by the number of electrons at the terminal cofactor site. These electrons are "created" ("destroyed") from the kinetic network when they move into (out of) their electron reservoir. For example, the "refilling" process by which $D^=$ donates its electrons to B is modelled by connecting states like

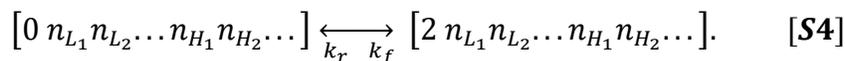

$$[0\ n_{L_1} n_{L_2}\ldots n_{H_1} n_{H_2}\ldots] \underset{k_r}{\overset{k_f}{\longleftrightarrow}} [2\ n_{L_1} n_{L_2}\ldots n_{H_1} n_{H_2}\ldots]. \qquad [S4]$$

The forward and reverse rate constants for this refill process are related as follows (reminisent of relations in interfacial electrochemistry (5))

$$k_f = k_r e^{-\Delta G/kT} = k_r\, exp\left(-\frac{2\mu_D + 2FE_B^{mid}}{k_B T}\right) \qquad [S5]$$

where $\mu_D$ is the (electro-) chemical potential of electrons in the D reservoir (this is adjusted to alter the driving force $\Delta G_{bifurc}$), and $E_B^{mid}$ is the midpoint (2-electron average) standard state reduction potential of the electron bifurcating species B (there is not a minus sign because of the definition of the reduction potential). Thus, the refilling process is modeled as a two-electron concerted transfer, where concerted means that no buildup of semiquinone (one-electron intermediate species) occurs during the refill process. This assumption is reasonable for the Q-cycle complexes, since the electron bifurcating quinone diffuses into the $Q_o$ site to initiate the electron bifurcation reaction (6). Since the semiquinone is unstable (7), little semiquinone is generated during this diffusion process. The reservoirs connected to the L and H branches are similar, but involve moving one electron at a time.

The kinetics equations were proceedurally generated and solved using a body of code written in Python for this study. The functions and all code used to generate the figures are available for download from GitHub (https://github.com/JYuly/EB_kinetics).

## Supplementary Figures

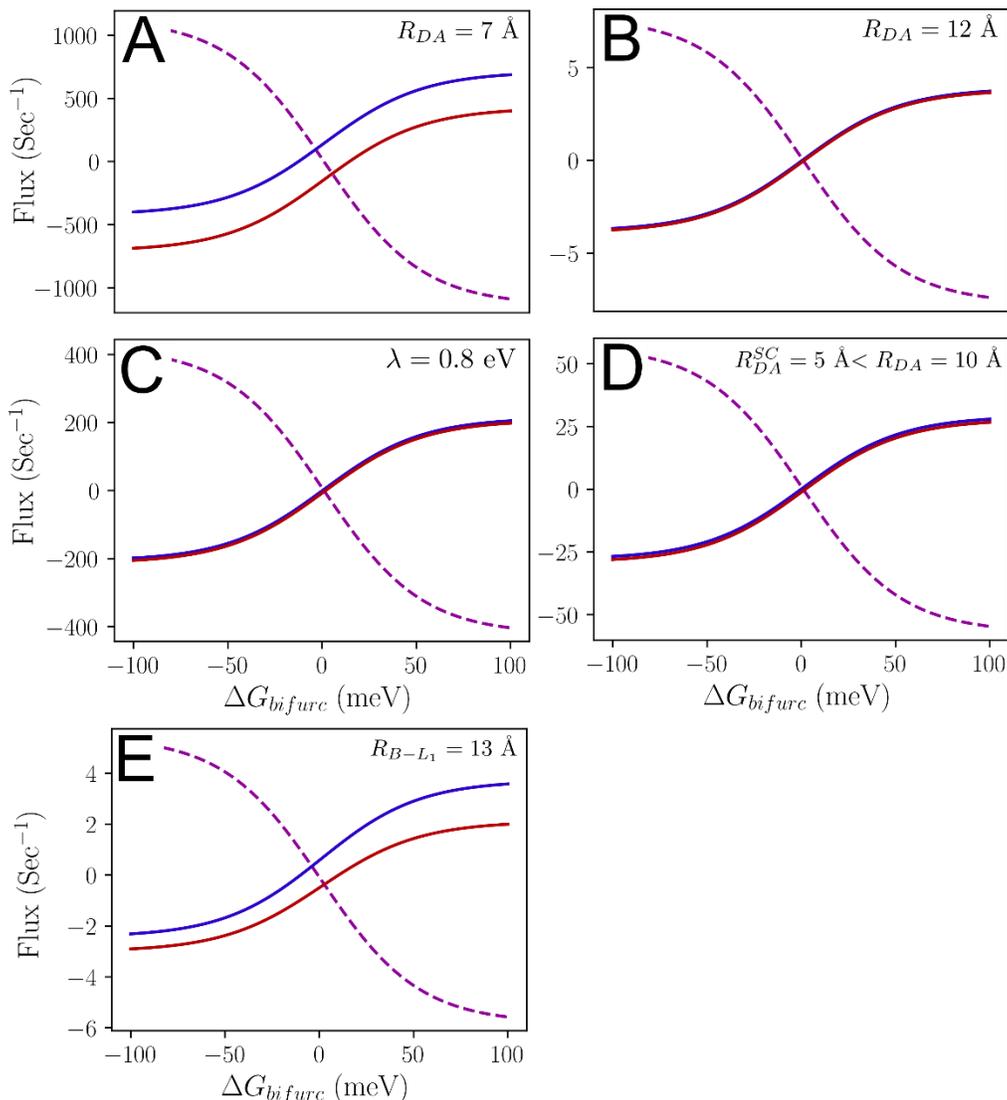

**Fig. S1 Dependence of electron bifurcation kinetics on model parameters.** The flux into each reservoir with various nearest neighbor tunneling distances $R_{DA}$ (A-B) and reorganization energy λ (C). Clear electron bifurcation and confurcation are preserved as these parameters are changed. Remarkably, when the short circuit rate constants are made to be larger than the productive rate constants (D), the EB-scheme retains highly efficient electron bifurcation and confurcation because of the Boltzmann suppression of short-circuits described in the text. The enhancement of the short circuits was induced by decreasing the effective tunneling distance ($R_{DA}^{SC}$) for the short-circuit electron tunneling rate constants ($R_{DA}^{SC} = 5$ Å) over the other electron transfer rate constants ($R_{DA} = 10$ Å for nearest neighbors, 20 Å for next neighbors, etc.). The turnover rate and short-circuit flux can change significantly if cofactor distances are outside of an ~ 5 Å window centered on the positions in the model of Fig. 3. For instance, if (E) the distance from D to $L_1$ ($R_{B-L_1}$) increased to 13 Å with all other distances fixed between nearest neighbors at 10 Å, the overall turnover rates are comparable to the case in (B), but with higher short-circuiting. Overall, the short-circuit flux when the EB-scheme is present never approaches the $10^5$/sec turnover indicated in Fig. 3B-C, which is found in the absence of the EB-scheme landscape.

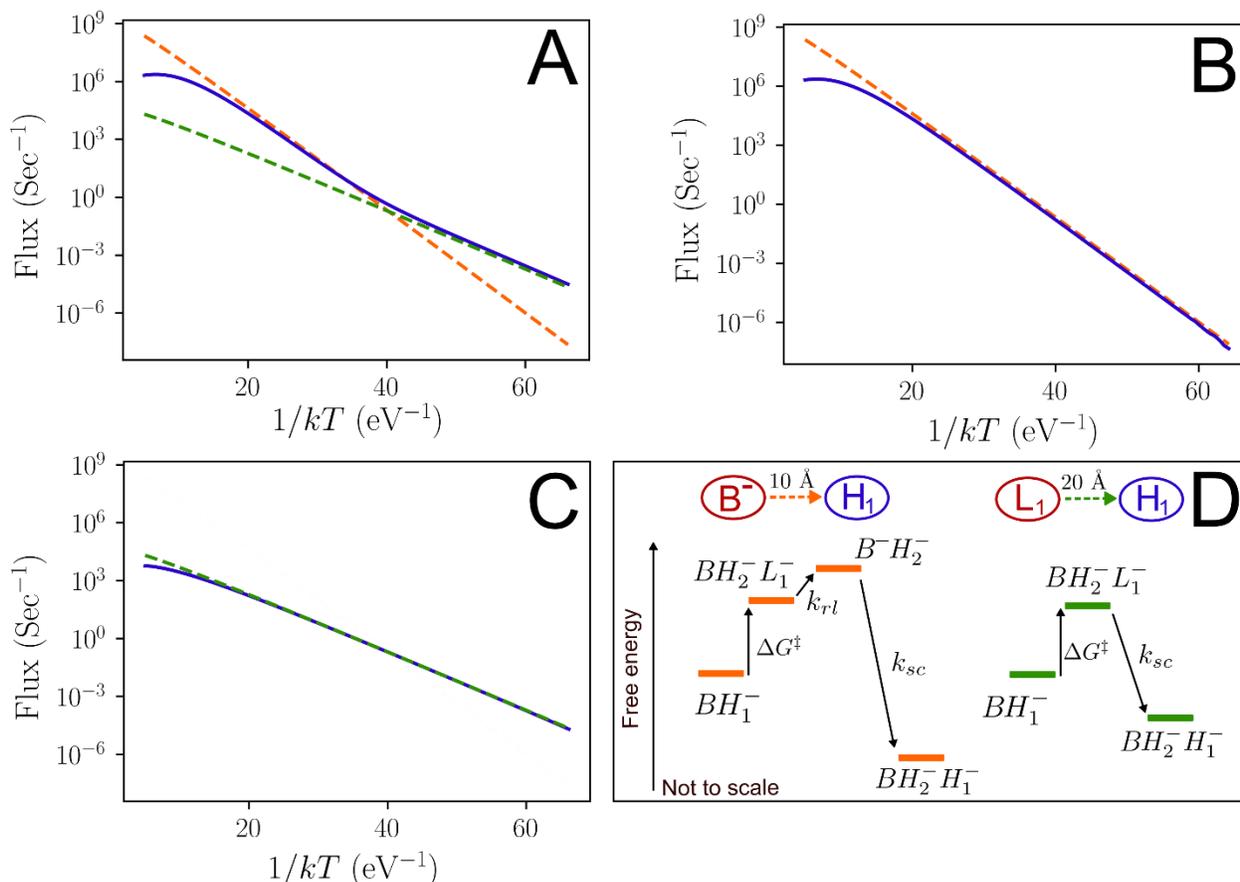

**Fig. S2 Short-circuiting in the EB-landscape is thermally activated.** The electron flux to the D, $A_H$, and $A_L$ reservoirs in the case with $\Delta G_{bifurc} = 0$ results from short-circuiting reactions. The calculated flux (A) falls into two linear regimes when plotted vs the inverse temperature (1/kT). The dashed lines correspond to approximate analytic expressions for the short-circuit flux that follows an Arrhenius-like expression (Eqs **S7** and **S8**), except at very high temperatures. These approximately linear regimes are determined by the dominating short-circuit process, which is verified by (B-C) setting all other short circuit rate constants equal to zero. The flux with only the dominant processes follows the same linear behavior. Kinetics diagrams (D) explain the important thermodynamic and kinetic features that determine the rates described by Eqs **S7** and **S8**. First, an electron and hole must move to the cofactor pair that will short circuit (with activation free energy $\Delta G^{\ddagger} = \Delta G_{electron} + \Delta G_{hole}$) to create an activated complex that can perform the sequence of kinetic steps required for short-circuiting. The rate limiting step in this process (rate constant

$k_{rl}$) is not necessarily the short-circuiting electron transfer itself (Fig. 1 C), such as the $L_1 \to H_1$ short-circuit. Given the parameters of the model, the $L_1 \to B^-$ short-circuit is identical in energy and rate constants to the $B^- \to H_1$ short-circuit, although the relevant redox states are different. Thus, in order to generate the orange dashed line, the contibutions from both the $L_1 \to B^-$ and $B^- \to H_1$ reactions were added together (their contributions are the same in this approximation). The analytical expressions for the short-circuiting are approximations; the simulated short-circuit fluxes based on full kinetic analysis were found to differ by as much as a factor of two from the flux calculated with Eq **S7**.

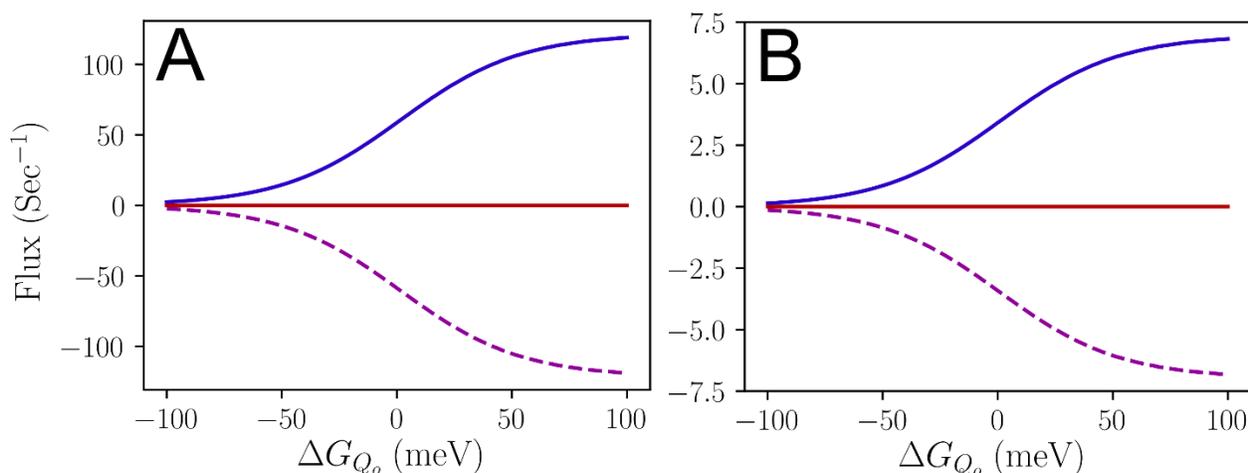

**Fig. S3 Steady-state behavior of the Q-cycle modeled with an inhibited low potential branch.** When (**a**) the $b_L$ to $b_H$ tunneling rate constant from the Q-cycle model of Fig. 4 (main text) is set to zero (analogous to $b_H$ knockouts (8, 9)), all of the electrons from $Q_o$ are routed to cyt *c* (solid blue), with a similar rate of quinone oxidation at the $Q_o$ site (dashed purple). The steady-state flux of electrons through the $Q_i$ site is zero, as electrons may not reach it from the $Q_o$ site or cyt *c*. This short-circuiting steady-state flux is not observed in experiments, which find short circuit fluxes about one order of magnitude lower (9). However, when (**b**) the effective distance $R_{eff}$ (Fig. 4) between the $Q_o$ site and the ISP is increased by 3 Å, the short-circuiting flux decreases by over an order of magnitude. This suggests that a conformational change (or even a change in the distance of closest approach between the ISP with the $Q_o$ site) of 3 Å or less can explain the kinetics of complex III with an inhibited low-potential branch. An even smaller conformational change may be needed, as proton-coupled electron transfer reactions can be very sensitive to donor-acceptor distance changes and the distances to proton donor/acceptor sites (10). While no experimental evidence for such subtle changes have been reported, the current modeling results suggest that subtle static or dynamic features may explain the observed low-potential branch inhibited kinetics. In any case, the role of the EB-scheme in preventing short-circuits is orders of magnitude larger than the observed difference between the inhibited and non-inhibited turnover rates for cytochrome *bc₁*.

## Supplementary Text

### Description of short-circuiting as an activated process:

For the landscape and model bifurcating landscape employed to produce Fig. 3, the flux of electrons into the mid-potential D reservoir is nearly zero when $\Delta G_{bifurc} = 0$. So in the case of Fig. 3, the short-circuiting steady-state turnover is the flux into the high-potential (blue) $A_H$ reservoir, since short-circuiting electrons do not flow into D. This is not always true. It is possible that short-circuiting could push electrons into the D reservoir, or out of the D reservoir and into the $A_H$ reservoir (this possibility is well known, see (8) and (11)).

The two linear regimes in Fig. 3 F suggest thermally activated short-circuits, each linear regime corresponding to a dominating process with activation energy proportional to the slope of each branch. This reasoning is dissected in Fig. S2 D. The two linear regimes are well approximated (to within a factor of ~ 2) using

$$flux \sim k_{rl}\, e^{-\Delta G^{\ddagger}/k_B T}, \qquad [S6]$$

where $k_{rl}$ is the rate-limiting rate constant in the sequence of steps required to short-circuit (Fig. S2 D). The rate-limiting $k_{rl}$ is not necessarily the rate constant for the short-circuit transfers $k_{SC}$ themselves (Fig. 1 C), although $k_{rl} = k_{SC}$ for the L$_1$ → H$_1$ short-circuit. Note that all rate constants employed in the model are temperature dependent (Eqs **S3** and **S5**), so the exponential term in **S6** is not the only source of temperature dependence.

The $\Delta G^{\ddagger}$ term is the free energy required to bring the system into the precursor state wherein the process described by $k_{rl}$ may occur. This contribution to the activation energy is a major contributing factor to the short-circuit insulating effect of the EB-scheme landscape, and the steep temperature dependence shown in Fig. 3 of the main text. The other major contribution to the activation free energy comes from the Frank-Condon factor in the non-adiabatic electron transfer rate expression (Eq **S3**). The energy $\Delta G^{\ddagger}$ required to bring the system into the precursor state is

$$\Delta G^{\ddagger} = \Delta G_{electron} + \Delta G_{hole}, \qquad [S7]$$

where $\Delta G_{electron}$ and $\Delta G_{hole}$ are the free energies to bring an electron and a hole to the the positions required to begin the short-circuit process. This sum of the activation free energies is the primary feature that suppresses short-circuiting over productive electron transfers.

The above kinetic reasoning was tested by setting all short-circuit rate constants to zero, except for the ones predicted by Eq **S6** and **S7** to dominate at a certain temperature. Then the analytical expressions (Eq **S6** and **S7)** were compared with results from the full kinetics described above. The results of this comparison appear in Fig. S2 B-C.

### Discussion of L-branch inhibition in complex III:

Experiments indicate that the steady-state turnover rate for cytochrome *bc*$_1$ drops by about a factor of 30 in the presence of antimycin and/or *b*-heme cofactor knockout mutants, both of which effectively inhibit electron flux through the low-potential branch (9). Our model does not exhibit this feature (Fig. 4 A, main text), which suggests that some aspects of the cytochrome *bc*$_1$ complex may not be captured by the model. This finding is not surprising given the simplicity of the model (Fig. 4 A, main text). However, antimycin-induced slowing does not abrogate the main conclusion of this study, namely that the EB-scheme accounts for most of the short-circuit insulation in the Q-cycle, as discussed below. This Supplementary Material explores the possible mechanisms that might account for the decreased turnover rates with a compromised L-branch. We argue that gating mechanisms are not necessary to resolve this mystery, in particular that subtle conformational changes and/or electrostatic interactions could be sufficient to explain decreased turnover rates. More importantly, we argue that if gating mechanisms underpin the change in turnover rates with a compromised L-branch, their contributions to the efficient partitioning of electrons to the high and low potential branches is minor compared to the powerful influence of the EB-landscape on the kinetics (this argument is put forward in the main text).

Since the dependence of electron transfer (or PCET) rates is exponential with respect to distance, free energy, and reorganization energy (2, 12, 13), small changes in these quantities could also generate the short-circuit suppression found in the L-branch inhibited turnover rates. Experimentally determining the molecular roots of this slightly more than one order of magnitude effect presents significant challenge, including whether the cause is a subtle conformational or dynamic change ~ 3 Å.

If an inhibited L-branch induces a conformational change of 3 Å or less at the $Q_o$ site, this would explain the kinetics of complex III with an inhibited low-potential branch (Fig. S3). Indeed, an even smaller conformational change (perhaps involving the distance to proton acceptors) may be needed to account for the experimental observation, as proton-coupled electron transfer reactions are very sensitive to distance changes (10). To the best of our knowledge, experiments have not demonstrated the absence of conformational changes to within a tolerance of 3 Å between an inhibited L-branch and the $Q_o$/Rieske FeS charge transfer precursor complex. This is not surprising, as the $Q_o$ site with quinone bound has not been crystalized, and conformational motion is known to be involved in the transfer from $Q_o$ to the Rieske FeS protein on the high-potential branch. Thus, the average distance between the Rieske center and the $Q_o$ site need not change due to low-potential branch inhibition for a conformational signal to reach the $Q_o$ site. Instead, only the distance of closest approach needs to change by about 3 Å. A dynamical effect of this kind would explain the L-branch inhibited kinetics, while maintaining average structural changes that are less than 3 Å. Even if the $Q_o$ site were known, the nature of the $Q_o$/Rieske charge-transfer precursor complex would need to be determined in order to exclude subtle conformational signaling effects on the scale of 3 Å. Such subtle effects on this length-scale should be dismissed only with care.

The crystal structures with stigmatellin or myxothiazol ($Q_o$ site inhibitors) could prevent subtle conformational changes induced by low-potential branch inhibition, which suggests that conformational changes (or lack thereof) for such structures must be interpreted with caution (14), and no structure has been observed with ubiquinone bound at the $Q_o$ site(s). Indeed, the stigmatellin and myxothiazol binding sites are different (15). It was suggested that there are, in fact, two binding sites for quinone that may correspond to $Q_o$ sites (11), and/or that quinone may move within the $Q_o$ site (16), so subtle conformational motion induced by low potential branch inhibitors may only influence one of the two sites (specifically, the site contributing the most to the turnover rates), explaining the order of magnitude decrease in turnover rates associated with L-branch inhibition. Indeed, one may ask whether or not conformational changes on this scale may be meaningfully ruled out based on static structural studies, as thermal motion may induce subtle changes on this scale that will not be captured in a static structure (17).

EPR studies have been carried out on the $Q_o$ site (18). However, when the Rieske center is oxidized and the $Q_o$ site contains a fully reduced hydroquinone, EPR experiments are insensitive to the distance between the Rieske center and the $Q_o$ site, as hydroquinone and oxidized FeS clusters are EPR silent. Specifically, the observed FeS/SQ$_o$ spin-coupled states may not be correlated with the conformational changes discussed here, as these changes need only appear when quinol occupies the $Q_o$ site and the Rieske FeS center is oxidized. In summary, it is difficult to rule out a ~ 3 Å conformational signal between the inhibited L-branch and the $Q_o \rightarrow$ Rieske FeS reaction (the rate limiting step of cytochrome $bc_1$ turnover (6)), as there are many ways for subtle conformational coupling to be transmitted.

In addition to a conformational coupling, it is also possible than an electrostatic interaction may be transmitted from an inhibited L-branch to the $Q_o$ site. For instance, if heme $b_L$ is reduced (i.e., if the L-branch is inhibited by antimycin or heme $b_H$ knockouts (9)), it will carry a negative charge, which may stabilize quinol relative to the anionic semiquinone at the $Q_o$ site. This would slow the overall rate-limiting proton-coupled electron transfer ($Q_o \rightarrow$ Rieske FeS), as it would be further uphill. This might explain the slowdown with an inhibited L-branch, but would not be a gating mechanism since the primary effect is on the productive $Q_o \rightarrow$ Rieske FeS proton-coupled electron transfer rate constant, and the effect would only persist with an inhibited L-branch. As well, this electrostatic interaction would not necessarily change the steady state semiquinone concentration at the $Q_o$ site, since a reduced $b_L$ heme also blocks the second electron from rapidly tunneling away from semiquinone on the L-branch, a process that normally destabilizes the semiquinone in the absence of a compromised L-branch (6). Thus, the destabilizing effect of the electrostatic interaction could compensate for the stabilizing effect of preventing semiquinone oxidation by the L-branch. This analysis suggests that the rate limiting $Q_o \rightarrow$ Rieske FeS electron transfer could be slowed by electrostatic interactions with a reduced $b_L$ heme, and that this slowing might explain a one-order of magnitude decrease in the turnover rate when the L-branch is compromised.

The considerations described in this Supplementary Information suggest that subtle static, dynamic, and electrostatic features (which are not necessarily gating mechanisms) may be responsible for the observed L-branch inhibited turnover rates of the cytochrome $bc_1$ enzyme, regardless of whether they fit into previously proposed gating schemes. In any case, the influence of the EB-scheme in disabling short-circuiting reactions is several orders of magnitude larger than the observed difference between the inhibited and non-inhibited turnover rates of cytochrome $bc_1$. As such, the EB-scheme provides at least the lion's share of the short-circuit insulation in the Q-cycle, especially under normal operating conditions.

**Supplementary references**